\documentclass[%
 aip,
 amsmath,amssymb,
 reprint,%
]{revtex4-1}

\usepackage{graphicx}
\usepackage{dcolumn}
\usepackage{bm}

\usepackage[utf8]{inputenc}
\usepackage[T1]{fontenc}
\usepackage{mathptmx}
\usepackage{etoolbox}
\usepackage{color}
\usepackage{url}

\makeatletter
\def\@email#1#2{%
 \endgroup
 \patchcmd{\titleblock@produce}
  {\frontmatter@RRAPformat}
  {\frontmatter@RRAPformat{\produce@RRAP{*#1\href{mailto:#2}{#2}}}\frontmatter@RRAPformat}
  {}{}
}%
\makeatother

\begin{document}

\preprint{AIP/123-QED}

\title[Spherical to Cartesian Coordinates Transformation for Solid Harmonics Revisited: Construction of the Hartree Potential]{Spherical to Cartesian Coordinates Transformation for Solid Harmonics Revisited: Construction of the Hartree Potential}
\author{Chiara Ribaldone}
\email{chiara.ribaldone@unito.it}
\author{Jacques Kontak Desmarais}
\email{jacqueskontak.desmarais@unito.it}
\affiliation{Dipartimento di Chimica, University of Torino, via Giuria 5, 10125, Torino, Italy}

\date{\today}

\begin{abstract}
Spherical Harmonic Gaussian type orbitals and Slater functions can be expressed using spherical coordinates or a linear combinations of the appropriate Cartesian functions. General expressions for the transformation coefficients between the two representations are provided. Values for the transformation coefficients are tabulated up to the quantum number $\ell = 10$. The formula is applied to construct the Hartree potential by an arbitrary-order multipole expansion.
\end{abstract}

\maketitle

\section{Introduction}

Harmonics are obtained as a solution to Laplace's equation and have a wide range of physical applications, for instance, in calculating the Coulomb potential,\cite{buehler1951bipolar} and in the representation of electromagnetic and gravitation fields \cite{takahashi2014small,carrascal1991vector} as well as acoustic modes in spherical coordinates.\cite{kumar2016near}  In many-body physics of extended systems, harmonics are employed for a multipole expansion of the Hartree potential.\cite{saunders1992electrostatic,watson2008linear,white1994derivation,kudin2000linear,orange_book} With reference to a given point $\bm{r} = (x, y, z)$ in a three dimensional space, unnormalized complex solid harmonics are defined as \cite{hobson1931theory,caola_fun}
\begin{equation}\label{eq1}
	Y_\ell^m(\bm{r}) = r^\ell \, P^{ \vert m \vert}_\ell(\cos\theta) \, e^{im\varphi}
\end{equation}
where $r = \lVert \bm{r} \rVert$ is the norm of the position vector, while $(\bm{r}, \theta, \varphi)$ are the spherical polar coordinates, and $P_\ell^{ \vert m \vert}(\cos\theta)$ is the associated Legendre function\cite{bell}
\begin{equation}\label{gingembre}
P^m_\ell(x)=\frac{1}{2^\ell \ell!} (1-x^2)^{m/2} \,\frac{d^{\ell+m}}{dx^{\ell+m}} (x^2-1)^\ell
\end{equation}
which we employ for quantum numbers $\ell \ge 0$ and $-\ell \le m \le \ell$.

In electronic structure theory, the spherical harmonics appear in the analytical solution of the Schr{\"o}dinger and Dirac equations for the Hydrogen atom. A success of quantum chemistry is the expansion of the many-body wavefunction into basis functions that resemble those of Hydrogen. The workhorse of quantum chemical calculations, the Gaussian-type orbitals (GTOs) of Boys and McWeeny,\cite{boys1950electronic,mcweeny1950gaussian} as well as Slater's functions,\cite{slater1932analytic} both employ spherical harmonics as a key component. 

The characteristic property of the spherical harmonic Gaussian-type orbitals (SGTOs) and Slater functions is that they remain eigenfunctions of the total electronic angular momentum operator $L^2$ and its $z$-component $L_z$, which is advantageous for interpretation of the calculation in terms of conventional chemical intuition. The many-electron integral algorithms, on the other hand, are most often developed in Cartesian coordinates.\cite{gill1994molecular,dupuis1976evaluation,helgaker1995gaussian,saunders1983molecular} Examples are the algorithms of McMurchie and Davidson,\cite{mcmurchie1978one} and of Obara and Saika,\cite{obara1986efficient,obara1988general} as well as their refinements.\cite{head1988method,gill1991prism,johnson1991exact,cisneros1993improved,gill1989efficient} Direct use of spherical harmonics in such algorithms leads to integrals which cannot be exactly factorized in terms of Cartesian products. Though, it remains advantageous to work with SGTOs, so as to reduce the amount of basis functions in the calculation, especially when employing high quantum numbers. For instance, at $\ell=2$, there are 6 distinct Cartesian Gaussian-type orbitals (CGTOs), but only 5 SGTOs. At quantum number $\ell=10$, there are 66 unique CGTOs, but merely 21 SGTOs.\cite{notebin} With SGTOs, an important, more recent development involves the use of resolution of the identity procedures.\cite{ahlrichs2004efficient,giese2008contracted,peels2020fast}

It is possible to work with the reduced SGTO $S_{\ell}^{m}(\bm{r})$ basis, while taking advantage of the convenience of CGTOs $G_{tuv}(\bm{r})$ for integral evaluation, through an expansion
\begin{equation}\label{eqexpand}
	S_{\ell}^{m}(\bm{r}) = \sum_{tuv}^\ell C_{tuv}^{\ell m} \, G_{tuv}(\bm{r}) \qquad  \text{ where } \quad t + u + v = \ell
\end{equation}
in which $S_{\ell}^{m}(\bm{r})= Y_\ell^m(\bm{r}) \, e^{-\alpha r^2}$ is a complex solid and unnormalized SGTO,  and:
\begin{equation}\label{cgto}
G_{tuv}(\bm{r})=x^t y^u z^v e^{-\alpha r^2} 
\end{equation}
is an (exactly factorizable) unnormalized CGTO, while $C_{tuv}^{\ell m}$ are linear coefficients. Analogous formulae may be derived for the Slater functions by a replacement of the radial functions $e^{-\alpha r^2}$ in favour of $e^{-\alpha r}$. 

The problem may be further simplified by considering that the harmonics satisfy $[Y_\ell^{m}(\bm{r})]^* =(-1)^m \, Y_\ell^{-m}(\bm{r})$.\cite{bell} Thus, we may only consider positive values of $m$. This also implies that the real functions
\begin{equation}\label{xlm}
    X_\ell^{m}(\bm{r}) = \begin{cases}
        \Re \left[ Y_\ell^{\vert m \vert}(\bm{r}) \right] & \quad m \ge 0 \\
        \\
        \Im \left[ Y_\ell^{\vert m \vert}(\bm{r}) \right] & \quad m < 0
    \end{cases}
\end{equation}
span the same space as the $Y_\ell^m(\bm{r})$, which means that the calculation may be formulated entirely in real algebra. The resulting basis functions $R_{\ell}^{m}(\bm{r})= X_\ell^m(\bm{r}) \, e^{-\alpha r^2}$ are conventionally known as real solid SGTOs. Unlike the complex $S_{\ell}^{m}(\bm{r})$ functions, the real $R_{\ell}^{m}(\bm{r})$ are not eigenfunctions of $L_z$.

Schlegel and Frisch proposed an analytical formula to calculate the coefficients $C_{tuv}^{\ell m}$ of Eq. \eqref{eqexpand}.\cite{frisch_1994} Their analysis, however, has significant limitations.\cite{noteerror} Refs. \onlinecite{caola_fun} and \onlinecite{helgaker2013molecular} provided separate and useful formulae for the representation of harmonics in Cartesian coordinates, but not a closed-form expression for the coefficients $C_{tuv}^{\ell m}$. Here, we provide the general formula for the coefficients $C_{tuv}^{\ell m}$ permitting a transformation from spherical to Cartesian forms, as well as explicit values of the transformation coefficients up to $\ell = 10$.

Our interest is in introducing CGTO based methods in the \textsc{Crystal} code, which traditionally employs a basis of real solid SGTOs to evaluate electron repulsion integrals for molecules as well as $d=1,2,3$ dimensional periodic systems.\cite{erba2022crystal23,desmarais2018generalization} The strategy includes a particularly efficient procedure for electrostatic summations in extended systems, employing the Ewald potential with a model density of distributed point multipoles.\cite{saunders1992electrostatic} Analytical gradients of the integrals are also available.\cite{desmarais2023efficient,doll2001analytical,doll2001implementation,doll2006analytical} Nonetheless, our results are transferrable to other types of calculations that employ spherical harmonics as a central component.

\section{Theory}

\subsection{Coefficients for complex harmonics}

Using Rodrigues' formula (page 48 of Ref. \onlinecite{bell}) for the associated Legendre function, we write
\begin{eqnarray}\label{eq2}
	P_\ell^{\lvert m \rvert} (\cos\theta) &=& \frac{1}{2^\ell \, \ell!}\left[1 - \left(\cos\theta\right)^2\right]^{\lvert m \rvert / 2} \, \nonumber \\
    &\times&\frac{d^{\ell+\lvert m \rvert}}{d(\cos\theta)^{\ell+\lvert m \rvert}} \left[\left(\cos\theta\right)^2 - 1\right]^\ell
\end{eqnarray}

Inserting this formula for the unnormalized complex solid harmonics, we rewrite Eq. \eqref{eq1} as
\begin{eqnarray}\label{eq3}
	Y_\ell^{\lvert m \rvert}(\bm{r}) &=& \frac{r^\ell}{2^\ell \, \ell!} \,\, e^{i\lvert m \rvert\varphi} \left[1 - \left(\cos\theta\right)^2\right]^{\lvert m \rvert / 2} \, \nonumber \\
    &\times& \frac{d^{\ell+\lvert m \rvert}}{d(\cos\theta)^{\ell+\lvert m \rvert}} \left[\left(\cos\theta\right)^2 - 1\right]^\ell 
\end{eqnarray}

The set of Cartesian coordinates can then be retrieved from the spherical system of coordinates
\begin{equation}\label{eq4}
    x = r \, \sin\theta \, \cos\varphi \qquad  y = r \, \sin\theta \, \sin\varphi \qquad  z = r \, \cos\theta
\end{equation}
where $r \in [0, \infty)$, $\theta \in [0,\pi]$ and $\varphi \in [0, 2\pi)$. These transformations can be used in the right hand side of Eq. \eqref{eq3} to express the solid harmonic functions in Cartesian coordinates. First of all, the exponential term on the left hand side of Eq. \eqref{eq3} can be reformulated as
\begin{equation}\label{eq5}
e^{i\lvert m \rvert\varphi} = \left( r\sin\theta \right)^{-\lvert m \rvert} (x + iy )^{\lvert m \rvert} = \left( r^2 - z^2 \right)^{-\lvert m \rvert/2} (x + iy )^{\lvert m \rvert}
\end{equation}

Using the previous Eq. \eqref{eq5}, together with the last transformation in Eq. \eqref{eq4}, the unnormalized complex solid harmonics of Eq. \eqref{eq3} can be written, as functions of the Cartesian coordinates, in the following way
\begin{equation}\label{eq6}
	\begin{aligned}
	& Y_\ell^{\lvert m \rvert}(\bm{r}) 
	 = \frac{r^{\ell - \lvert m \rvert}}{2^\ell \, \ell!} \, (x + iy )^{\lvert m \rvert} \, \frac{d^{\ell+\lvert m \rvert}}{d(z/r)^{\ell+\lvert m \rvert}} \left[\left(\frac{z}{r}\right)^2 - 1\right]^\ell \\
	& = \frac{r^{\ell - \lvert m \rvert}}{2^\ell \, \ell!} \, (x + iy )^{\lvert m \rvert} \, \sum_{k = 0}^\ell \binom{\ell}{k} (-1)^{k} \frac{d^{\ell+\lvert m \rvert}}{d(z/r)^{\ell+\lvert m \rvert}} \left(\frac{z}{r}\right)^{2\ell - 2k}
    \end{aligned}
\end{equation}
where the binomial theorem has been used in the last identity to conveniently rewrite the polynomial on which the derivative acts. In order to evaluate the derivative in the last expression, the following basic rule can be verified
\begin{equation}\label{derivative_rule}
    \left(\frac{\partial}{\partial x}\right)^n (x + a)^m = \frac{m!}{(m-n)!} \, (x+a)^{m-n} \, H[m - n]
\end{equation}
where $H[n]$ is the Heaviside step function in the $H[0]=1$ convention. Therefore, Eq. \eqref{eq6} can be expressed more explicitly as 
\begin{align}\label{eq7}
    &Y_\ell^{\lvert m \rvert}(\bm{r}) = \frac{1}{2^\ell \, \ell!} \, (x + iy )^{\lvert m \rvert} \, \nonumber \\
    &\times \sum_{k = 0}^{\sigma_{\ell \lvert m \rvert}} \binom{\ell}{k} \, \frac{(-1)^{k} \, (2\ell - 2k)!}{(\ell - \lvert m \rvert - 2k)!} \,\, r^{2k} \, {z}^{\ell - \lvert m \rvert - 2k}
\end{align}
where $\sigma_{\ell \lvert m \rvert}=\lfloor (\ell - \lvert m \rvert)/2 \rfloor$ and the upper limit of the summation is fixed to $k \le \sigma_{\ell \lvert m \rvert}$ (with $\lfloor ... \rfloor$ denoting the round down operation) because the derivative in Eq. \eqref{eq6} vanishes if its order exceeds the degree of the monomial $(z/r)$. At this point, using the relation
\begin{equation}\label{eq9}
    r^2 = x^2 + y^2 + z^2
\end{equation} 
Eq. \eqref{eq7} can reformulated, in terms of Cartesian coordinates, as
\begin{eqnarray}\label{eq10}
    Y_\ell^{\lvert m \rvert}(\bm{r}) = \frac{1}{2^\ell \, \ell!} \, (x + iy )^{\lvert m \rvert} \, \sum_{k = 0}^{\sigma_{\ell \lvert m \rvert}} \binom{\ell}{k} \, \frac{(-1)^{k} \, (2\ell - 2k)!}{(\ell - \lvert m \rvert - 2k)!} \,\,  \nonumber \\
    \times \, (x^2 + y^2 + z^2)^{k} \, {z}^{\ell - \lvert m \rvert - 2k}
\end{eqnarray}

Eq. \eqref{eq10} provides a representation of solid harmonics as a sum of polynomials of the Cartesian coordinates. They can be converted to a sum of monomials of the coordinates by considering the expansion
\begin{equation}\label{eq11}
    Y_\ell^{\lvert m \rvert}(\bm{r}) = \sum_{tuv}^{\ell} C_{tuv}^{\ell \lvert m \rvert} \,\, x^t y^u z^v \qquad  \text{ where } \quad t + u + v = \ell
\end{equation}
and by differentiating both sides of Eq. \eqref{eq11} with respect to the $(x,y,z)$ variables and then evaluating the derivatives at $x = y = z = 0$.
To this end, equation \eqref{eq10} can be reformulated using the binomial theorem leading to
\begin{align}\label{eq13}
    &\sum_{tuv}^{\ell} C_{tuv}^{\ell \lvert m \rvert} \,\, x^t y^u z^v \nonumber \\
    & = \frac{1}{2^\ell \, \ell!} \, \sum_{s = 0}^{\lvert m \rvert} \sum_{k = 0}^{\sigma_{\ell \lvert m \rvert}} \binom{\lvert m \rvert}{s} \binom{\ell}{k} \, \frac{(-1)^{k + (\lvert m \rvert - s)/2} \, (2\ell - 2k)!}{(\ell - \lvert m \rvert - 2k)!} \,\,  \nonumber \\
    & \quad \times \sum_{p = 0}^k \sum_{q = 0}^p \binom{k}{p} \binom{p}{q} \, x^{s + 2q} \, y^{\lvert m \rvert - s + 2p - 2q} \, {z}^{\ell - \lvert m \rvert - 2p}
\end{align}
with the restriction $t + u + v = \ell$ for the summation in the left hand side. \\
At this point, by applying the derivatives $\partial_x^\tau \, \partial_y^\mu \, \partial_z^\nu$ on both sides of Eq. \eqref{eq13}, and evaluating the derivatives at $x = y = z = 0$, the analytic expression for the expansion coefficients $C_{tuv}^{\ell \lvert m \rvert}$ can be easily found. Indeed, taking the derivatives on the left hand side of Eq. \eqref{eq13} by means of Eq. \eqref{derivative_rule}, gives the result
\begin{align}\label{eq14}
    &\left . \partial_x^\tau \, \partial_y^\mu \, \partial_z^\nu \left[ \, \sum_{tuv}^{\ell} C_{tuv}^{\ell \lvert m \rvert} \,\, x^t y^u z^v \right] \right\rvert_{x,y,z = 0}  \nonumber \\
    &= \tau! \, \mu! \, \nu! \, C_{\tau\mu\nu}^{\ell \lvert m \rvert} =  \left. \partial_x^\tau \, \partial_y^\mu \, \partial_z^\nu  \left[ Y_\ell^{\lvert m \rvert}(\bm{r})\right]\right\rvert_{x,y,z = 0}
\end{align}
where the order of the derivatives must respect the condition $\tau + \mu + \nu = \ell$ imposed for the order of the monomial appearing in the summation, as defined in the expansion of Eq. \eqref{eq11}.

\begin{widetext}
{
\setlength{\tabcolsep}{10.2pt}
\renewcommand{\arraystretch}{1.28}
\begin{table}\small 
    \centering
    \begin{tabular}{c|rrrrrr}
         \toprule
         $(\ell,m)$ & \multicolumn{6}{c}{Non zero $D_{tuv}^{\ell m}$} \\
         \hline
         $(0,  0)$ & 1 $(0, 0, 0)$ & & & & & \\
         \hline
         $(1, -1)$ & 1 $(0, 1, 0)$ & & & & & \\
         $(1,  0)$ & 1 $(0, 0, 1)$ & & & & & \\
         $(1,  1)$ & 1 $(1, 0, 0)$ & & & & & \\
         \hline
         $(2, -2)$ & 6 $(1, 1, 0)$ & & & & & \\
         $(2, -1)$ & 3 $(0, 1, 1)$ & & & & & \\
         $(2,  0)$ & 1 $(0, 0, 2)$ & -1/2 $(0, 2, 0)$ & -1/2 $(2, 0, 0)$ & & & \\
         $(2,  1)$ & 3 $(1, 0, 1)$ & & & & & \\
         $(2,  2)$ & -3 $(0, 2, 0)$ & 3 $(2, 0, 0)$ & & & & \\
         \hline
         $(3, -3)$ & -15 $(0, 3, 0)$ & 45 $(2, 1, 0)$ & & & & \\
         $(3, -2)$ & 30 $(1, 1, 1)$ & & & & \\
         $(3, -1)$ & 6 $(0, 1, 2)$ & -3/2 $(0, 3, 0)$ & -3/2 $(2, 1, 0)$ & & & \\
         $(3,  0)$ & 1 $(0, 0, 3)$ & -3/2 $(0, 2, 1)$ & -3/2 $(2, 0, 1)$ & & & \\
         $(3,  1)$ & 6 $(1, 0, 2)$ & -3/2 $(1, 2, 0)$ & -3/2 $(3, 0, 0)$ & & & \\
         $(3,  2)$ & -15 $(0, 2, 1)$ & 15 $(2, 0, 1)$ & & & & \\
         $(3,  3)$ & -45 $(1, 2, 0)$ & 15 $(3, 0, 0)$ & & & & \\
         \hline
         $(4, -4)$ & -420 $(1, 3, 0)$ & 420 $(3, 1, 0)$ & & & & \\
         $(4, -3)$ & -105 $(0, 3, 1)$ & 315 $(2, 1, 1)$ & & & & \\
         $(4, -2)$ & 90 $(1, 1, 2)$ & -15 $(1, 3, 0)$ & -15 $(3, 1, 0)$ & & & \\
         $(4, -1)$ & 10 $(0, 1, 3)$ & -15/2 $(0, 3, 1)$ & -15/2 $(2, 1, 1)$ & & & \\
         $(4,  0)$ & 1 $(0, 0, 4)$ & -3 $(0, 2, 2)$ & 3/8 $(0, 4, 0)$ & -3 $(2, 0, 2)$ & 3/4 $(2, 2, 0)$ & 3/8 $(4, 0, 0)$ \\
         $(4,  1)$ & 10 $(1, 0, 3)$ & -15/2 $(1, 2, 1)$ & -15/2 $(3, 0, 1)$ & & & \\
         $(4,  2)$ & -45 $(0, 2, 2)$ & 15/2 $(0, 4, 0)$ & 45 $(2, 0, 2)$ & -15/2 $(4, 0, 0)$ & & \\
         $(4,  3)$ & -315 $(1, 2, 1)$ & 105 $(3, 0, 1)$ & & & & \\
         $(4,  4)$ & 105 $(0, 4, 0)$ & -630 $(2, 2, 0)$ & 105 $(4, 0, 0)$ & & & \\
         \toprule
    \end{tabular}
    \caption{Non zero $D_{tuv}^{\ell m}$ coefficients in the polynomial representation of the real solid harmonics up to $\ell=4$. The triplet of integers in brackets after each coefficient specifies the corresponding exponents $t,u,v$. For instance: $X_3^{3}=15 x^3-45xy^2$}
    \label{tab:clmtuv_l04}
\end{table}
}
\end{widetext}

On the other hand, taking the derivatives on the right hand side of Eq. \eqref{eq13} by means of Eq.  \eqref{derivative_rule} leads to the expression
\begin{widetext}
\begin{equation}\label{eq15}
	\partial_x^\tau \, \partial_y^\mu \, \partial_z^\nu \left. \left[ \, \sum_{p = 0}^k \sum_{q = 0}^p \binom{k}{p} \binom{p}{q} \, x^{s + 2q} \, y^{\lvert m \rvert - s + 2p - 2q} \, {z}^{\ell - \lvert m \rvert - 2p} \right] \right\rvert_{x,y,z = 0}
     = \tau! \, \mu! \, \nu! \, \sum_{p = 0}^k \sum_{q = 0}^p \binom{k}{p} \binom{p}{q} \, \delta^{\tau}_{s + 2q} \, \delta^\mu_{\lvert m \rvert - s + 2p - 2q} \, \delta^\nu_{\ell - \lvert m \rvert - 2p}
\end{equation}
\end{widetext}

The Kronecker delta functions in the previous expression can be rearranged as follows
\begin{align}\label{delta_relations}
	& \delta^{\tau}_{s + 2q} \, \delta^\mu_{\lvert m \rvert - s + 2p - 2q} \, \delta^\nu_{\ell - \lvert m \rvert - 2p} = \delta^{\tau}_{s + 2q} \, \delta^\mu_{\lvert m \rvert - s + 2p - 2q} \, \delta^\nu_{\ell - \mu - s - 2q} \nonumber \\
	& = \delta^{\tau}_{s + 2q} \, \delta^\mu_{\lvert m \rvert - s + 2p - 2q} \, \delta^\nu_{\ell - \mu - \tau} = \delta^{\tau}_{s + 2q} \, \delta^\mu_{\lvert m \rvert - s + 2p - 2q} \, \delta^{\tau + \mu + \nu}_{\ell}
\end{align}
where in the first equality the condition imposed by the second delta function,
\begin{equation}\label{eq16}
	\lvert m \rvert - s + 2p - 2q = \mu \qquad \to \qquad \lvert m \rvert + 2p = \mu + s + 2q
\end{equation}
has been applied on the indices of the third delta function, while in the second identity the condition established by the first delta function,
\vspace{-0.28cm}
\begin{equation}\label{eq17}
	s + 2q = \tau
\end{equation}
has been used to rewrite the indices of the third delta. Now, combining the results obtained in Eqs. \eqref{eq14} and \eqref{eq15} with the formula \eqref{eq13} yields the expression

\begin{widetext}
\begin{align}\label{eq18}
		 & \partial_x^\tau \, \partial_y^\mu \, \partial_z^\nu \left. \left[ Y_\ell^{\lvert m \rvert}(\bm{r}) \right] \right\rvert_{x,y,z = 0} = \notag \\
		& \partial_x^\tau \, \partial_y^\mu \, \partial_z^\nu \left. \left[ \frac{1}{2^\ell \, \ell!} \, \sum_{s = 0}^{\lvert m \rvert} \sum_{k = 0}^{\sigma_{\ell \lvert m \rvert}} \binom{\lvert m \rvert}{s} \binom{\ell}{k} \, \frac{(-1)^{k + (\lvert m \rvert - s)/2} \, (2\ell - 2k)!}{(\ell - \lvert m \rvert - 2k)!} \,\, \sum_{p = 0}^k \sum_{q = 0}^p \binom{k}{p} \binom{p}{q} \, x^{s + 2q} \, y^{\lvert m \rvert - s + 2p - 2q} \, {z}^{\ell - \lvert m \rvert - 2p} \right] \right\rvert_{x,y,z = 0} \notag \\
		& = \tau! \, \mu! \, \nu! \left[ \frac{1}{2^\ell \, \ell!} \, \sum_{s = 0}^{\lvert m \rvert} \sum_{k = 0}^{\sigma_{\ell \lvert m \rvert}} \sum_{p = 0}^k \sum_{q = 0}^p \binom{\ell}{k} \binom{k}{p} \binom{p}{q} \binom{\lvert m \rvert}{s} \, \frac{(-1)^{k + (\lvert m \rvert - s)/2} \, (2\ell - 2k)!}{(\ell - \lvert m \rvert - 2k)!} \,\,\, \delta^{\tau}_{s + 2q} \, \delta^\mu_{\lvert m \rvert - s + 2p - 2q} \right] \delta^{\tau + \mu + \nu}_{\ell} 
\end{align}
\end{widetext}

Considering the last line of Eq. \eqref{eq18}, the delta functions induce the following constraints on summation indices, as well as quantum numbers
\vspace{-0.20cm}
\begin{equation}\label{eq21}
    0 \le \tau - 2q \le \lvert m \rvert
\end{equation}
and
\begin{equation}\label{eq22}
    \frac{\tau + \mu - \lvert m \rvert}{2} \le k \qquad \quad \text{with} \quad \frac{\tau + \mu - \lvert m \rvert}{2} \in \mathbb{N}
\end{equation}

where $\mathbb{N}=0,1,2,\dots$ denotes the set of natural numbers. Eqs. \eqref{eq21} and \eqref{eq22} may be imposed onto Eq. \eqref{eq18} as follows
\begin{widetext}
\begin{align}\label{eqbrutta}
&  \partial_x^\tau \, \partial_y^\mu \, \partial_z^\nu \left. \left[  Y_\ell^{\lvert m \rvert}(\bm{r}) \right] \right\rvert_{x,y,z = 0} = \nonumber \\
&  \tau! \, \mu! \, \nu! \left[ \frac{M_{\tau\mu \lvert m \rvert }}{2^\ell \, \ell!} \, \sum_{k = 0}^{\sigma_{\ell \lvert m \rvert}} \sum_{q = 0}^{\eta_{\tau \mu \lvert m \rvert }}  \binom{\ell}{k} \binom{k}{\eta_{\tau \mu \lvert m \rvert } } \binom{ \eta_{\tau \mu  \lvert m \rvert }  }{q} \binom{\lvert m \rvert}{\tau - 2q} \, \frac{(-1)^{k + (\lvert m \rvert - \tau + 2q)/2} \, (2\ell - 2k)!}{(\ell - \lvert m \rvert - 2k)!} \,\, \Theta_{\tau\mu  \lvert m \rvert  k q} \right] \delta^{\tau + \mu + \nu}_{\ell}
\end{align}
\end{widetext}

in which the compact notation:

\begin{equation}
 M_{\tau\mu \lvert m \rvert } = \left[1-{\rm mod}(\tau + \mu - \lvert m \rvert,2) \right] \, H [\tau+\mu-\lvert m \rvert]
\end{equation}
\begin{equation}
    \eta_{\tau \mu \lvert m \rvert } \equiv \frac{\tau + \mu - \lvert m \rvert}{2} 
\end{equation}
and
\begin{equation}\Theta_{\tau\mu  \lvert m \rvert k q} = H [k-\eta_{\tau \mu \lvert m \rvert}] \ H [\tau - 2q] \ H [\lvert m \rvert - \tau +2q] 
\end{equation}
has been introduced, with ${\rm mod}(n,m)$ representing the modulus of division $n/m$, and $H[n]$ being the Heaviside step function in the $H[0]=1$ convention. Combining Eqs. \eqref{eq14} and \eqref{eqbrutta}, we finally obtain

\begin{widetext}
\begin{equation}\label{eq19}
	\begin{aligned}
	C_{tuv}^{\ell \lvert m \rvert} =
        \frac{M_{tu \lvert m \rvert }}{2^\ell \, \ell!} \, \sum_{k = 0}^{\sigma_{\ell \lvert m \rvert}}  \binom{\ell}{k} \binom{k}{\eta_{tu \lvert m \rvert }} \,\, \frac{(-1)^{k} \, (2\ell - 2k)!}{(\ell - \lvert m \rvert - 2k)!} \,\, \sum_{q = 0}^{\eta_{tu \lvert m \rvert }} \binom{\eta_{tu \lvert m \rvert }}{q} \binom{\lvert m \rvert}{t - 2q} \, (-1)^{(\lvert m \rvert - t + 2q)/2} \, \Theta_{tu  \lvert m \rvert k q} \, \delta^{t + u + v}_{\ell}
	\end{aligned}
\end{equation}
\end{widetext}

where the free indices $(\tau,\mu,\nu)$ have been renamed as $(t,u,v)$, respectively.
The most important difference of our Eq. \eqref{eq19} in comparison with the result of Ref. \onlinecite{frisch_1994} is that we obtained a round down operation on the upper limit $\sigma_{\ell \lvert m \rvert}=\lfloor (\ell - \lvert m \rvert)/2 \rfloor$  of the $k$ summation.

\subsection{Coefficients for real harmonics}

The coefficients $D_{tuv}^{\ell m}$ associated with the unnormalized real solid harmonics can then be derived from Eq. \eqref{xlm}. Indeed, taking the real (for $m \ge 0$) and imaginary (for $m < 0$) parts in Eq. \eqref{eq11}, leads to the equation
\begin{equation}\label{eq30}
    X_\ell^{m}(\bm{r}) = \sum_{tuv}^{\ell} D_{tuv}^{\ell m} \,\, x^t y^u z^v  \qquad \text{ where } \quad t + u + v = \ell
\end{equation}
with
\begin{equation}\label{eqD1}
    D_{tuv}^{\ell  m } = \begin{cases}
        \Re \left[ C_{tuv}^{\ell \lvert m \rvert} \right] & \quad m \ge 0 \\
        \\
        \Im \left[ C_{tuv}^{\ell \lvert m \rvert} \right] & \quad m < 0 \\
    \end{cases}
\end{equation}

Values of the $D_{tuv}^{\ell m}$ coefficients  calculated through Eqs. \eqref{eq19} and \eqref{eqD1} are listed in Tables \ref{tab:clmtuv_l04} and \ref{tab:clmtuv_l56}, up to the quantum number $\ell = 6$, while we provide the Fortran08 source code up to a user provided maximum value of the $\ell$ quantum number, as well as tabulated values up to $\ell = 10$, in the Supplementary Material.

 \begin{widetext}
{ 
\setlength{\tabcolsep}{10pt}
\renewcommand{\arraystretch}{1.28}
\begin{table}\small 
    \centering
    \begin{tabular}{c|rrrrrr}
         \toprule
         $(\ell,m)$ & \multicolumn{6}{c}{Non zero $D_{tuv}^{\ell m}$} \\
         \hline
         $(5, -5)$ & 945 $(0, 5, 0)$ & -9450 $(2, 3, 0)$ & 4725 $(4, 1, 0)$ & & \\
         $(5, -4)$ & -3780 $(1, 3, 1)$ & 3780 $(3, 1, 1)$ & & & & \\
         $(5, -3)$ & -420 $(0, 3, 2)$ & 105/2 $(0, 5, 0)$ & 1260 $(2, 1, 2)$ & -105 $(2, 3, 0)$ & -315/2 $(4, 1, 0)$ & \\
         $(5, -2)$ & 210 $(1, 1, 3)$ & -105 $(1, 3, 1)$ & -105 $(3, 1, 1)$ & & & \\
         $(5, -1)$ & 15 $(0, 1, 4)$ & -45/2 $(0, 3, 2)$ & 15/8 $(0, 5, 0)$ & -45/2 $(2, 1, 2)$ & 15/4 $(2, 3, 0)$ & 15/8 $(4, 1, 0)$ \\
         $(5,  0)$ & 1 $(0, 0, 5)$ & -5 $(0, 2, 3)$ & 15/8 $(0, 4, 1)$ & -5 $(2, 0, 3)$ & 15/4 $(2, 2, 1)$ & 15/8 $(4, 0, 1)$ \\
         $(5,  1)$ & 15 $(1, 0, 4)$ & -45/2 $(1, 2, 2)$ & 15/8 $(1, 4, 0)$ & -45/2 $(3, 0, 2)$ & 15/4 $(3, 2, 0)$ & 15/8 $(5, 0, 0)$ \\
         $(5,  2)$ & -105 $(0, 2, 3)$ & 105/2 $(0, 4, 1)$ & 105 $(2, 0, 3)$ & -105/2 $(4, 0, 1)$ & & \\
         $(5,  3)$ &  -1260 $(1, 2, 2)$ & 315/2 $(1, 4, 0)$ & 420 $(3, 0, 2)$ & 105 $(3, 2, 0)$ & -105/2 $(5, 0, 0)$ & \\
         $(5,  4)$ & 945 $(0, 4, 1)$ & -5670 $(2, 2, 1)$ & 945 $(4, 0, 1)$ & & & \\
         $(5,  5)$ & 4725 $(1, 4, 0)$ & -9450 $(3, 2, 0)$ & 945 $(5, 0, 0)$ & & & \\        
         \hline
         $(6, -6)$ & 62370 $(1, 5, 0)$ & -207900 $(3, 3, 0)$ & 62370 $(5, 1, 0)$ & & & \\
         $(6, -5)$ & 10395 $(0, 5, 1)$ & -103950 $(2, 3, 1)$ & 51975 $(4, 1, 1)$ & & & \\
         $(6, -4)$ & -18900 $(1, 3, 2)$ & 1890 $(1, 5, 0)$ & 18900 $(3, 1, 2)$ & -1890 $(5, 1, 0)$ & & \\
         $(6, -3)$ & -1260 $(0, 3, 3)$ & 945/2  $(0, 5, 1)$ & 3780 $(2, 1, 3)$ & -945 $(2, 3, 1)$ & -2835/2 $(4, 1, 1)$ & \\
         $(6, -2)$ & 420 $(1, 1, 4)$ & -420 $(1, 3, 2)$ & 105/4 $(1, 5, 0)$ & -420 $(3, 1, 2)$ & 105/2 $(3, 3, 0)$ & 105/4 $(5, 1, 0)$ \\
         $(6, -1)$ & 21 $(0, 1, 5)$ & -105/2 $(0, 3, 3)$ & 105/8 $(0, 5, 1)$ & -105/2 $(2, 1, 3)$ & 105/4 $(2, 3, 1)$ & 105/8 $(4, 1, 1)$ \\
         $(6,  0)$ & 1 $(0, 0, 6)$ & -15/2 $(0, 2, 4)$ & 45/8 $(0, 4, 2)$ & -5/16 $(0, 6, 0)$ & -15/2 $(2, 0, 4)$ & 45/4 $(2, 2, 2)$ \\
         & -15/16 $(2, 4, 0)$ & 45/8 $(4, 0, 2)$ & -15/16 $(4, 2, 0)$ & -5/16 $(6, 0, 0)$ & & \\
         $(6,  1)$ & 21 $(1, 0, 5)$ & -105/2 $(1, 2, 3)$ & 105/8 $(1, 4, 1)$ & -105/2 $(3, 0, 3)$ & 105/4 $(3, 2, 1)$ & 105/8 $(5, 0, 1)$ \\
         $(6,  2)$ & -210 $(0, 2, 4)$ & 210 $(0, 4, 2)$ & -105/8 $(0, 6, 0)$ & 210 $(2, 0, 4)$ & -105/8 $(2, 4, 0)$ & -210 $(4, 0, 2)$ \\
         & 105/8 $(4, 2, 0)$ & 105/8 $(6, 0, 0)$ & & & & \\
         $(6,  3)$ & -3780 $(1, 2, 3)$ & 2835/2 $(1, 4, 1)$ & 1260 $(3, 0, 3)$ & 945 $(3, 2, 1)$ & -945/2 $(5, 0, 1)$ & \\
         $(6,  4)$ & 4725 $(0, 4, 2)$ & -945/2 $(0, 6, 0)$ & -28350 $(2, 2, 2)$ & 4725/2 $(2, 4, 0)$ & 4725 $(4, 0, 2)$ & 4725/2 $(4, 2, 0)$ \\
         & -945/2 $(6, 0, 0)$ & & & & & \\
         $(6,  5)$ & 51975 $(1, 4, 1)$ & -103950 $(3, 2, 1)$ & 10395 $(5, 0, 1)$ & & & \\
         $(6,  6)$ & -10395 $(0, 6, 0)$ & 155925 $(2, 4, 0)$ & -155925 $(4, 2, 0)$ & 10395 $(6, 0, 0)$ & & \\
         \toprule
    \end{tabular}
    \caption{Same as Table \ref{tab:clmtuv_l04}, but for quantum number $\ell=5$ and $\ell=6 $.}
    \label{tab:clmtuv_l56}
\end{table}
}
\end{widetext}

\section{Construction of the Hartree Potential}

As an application of the new formula, Eq. \eqref{eq19}, we consider the construction of the Hartree potential $\Phi\left[ n\right]$ created by the electron density $n$ of molecular systems. This potential, evaluated at a point $\bm{r}$, external to a sphere centered at $\mathbf{R}$ which contains the whole charge distribution $n$, is given by the Neumann-Laplace expansion of $1/\rVert\bm{r}-\bm{r}'\lVert$ and reads
\begin{equation}\label{eqn:Phi}
\begin{aligned}
\Phi\left[ n\right] (\bm{r}) & = \int d^3 r' \ \frac{n(\bm{r}')}{\rVert\bm{r}-\bm{r}'\lVert} \\
& = \sum_{\ell=0}^\infty \sum_{m=-\ell}^{\ell} \eta^m_{\ell} \left[ n \right](\mathbf{R}) \ Z^m_{\ell} (\partial/\partial{\mathbf{R}}) \frac{1}{\rVert\bm{r}-\mathbf{R}\lVert}
\end{aligned}
\end{equation}
where $\eta^m_{\ell} \left[ n \right](\mathbf{R})$ are the multipole moments of $n$
\begin{equation}\label{eqn:mom}
\eta^m_{\ell} \left[ n \right](\mathbf{R})=\int d^3 r' \ n(\bm{r}') \, X^m_\ell (\bm{r}'-\mathbf{R})
\end{equation}
while $Z^m_{\ell}(\partial/\partial{\mathbf{R}})$ is the normalized real-spherical-gradient operator \cite{saunders1992electrostatic,hobson1931theory}
\begin{equation}
Z^m_{\ell}(\partial/\partial{\mathbf{R}})=\frac{(2-\delta_{m0}) [(\ell-|m|)!]}{[(2\ell-1)!!][(\ell+|m|)!]} \,\, X^m_\ell (\partial/\partial{\mathbf{R}})
\end{equation}
The upshot of Eq. \eqref{eqn:Phi} is that it has transformed the two-electron operator $1/\rVert\bm{r}-\bm{r}'\lVert$ to products of one-electron quantities, which can allow for the development of linear scaling algorithms.\cite{watson2008linear} Moreover, in extended systems, this formulation leads to quickly convergent lattice summations.\cite{saunders1992electrostatic,orange_book}

To evaluate the multipole moments $\eta^m_{\ell} \left[ n \right](\mathbf{R})$, we first introduce the real-solid spherical Gaussian atomic orbitals
\begin{equation}
\chi_\mu (\bm{r})= N_\mu \sum_{t u v}^{\ell} D^{\ell m}_{tuv} \ \prod_{q=x,y,z} (q-R_{\mu.q})^{i_q} \ e^{- \alpha_\mu (q-R_{\mu,q})^2} 
\end{equation}
with $\mu$ being a composite index for quantum numbers $\ell, m$ and atomic positions, while $N_\mu=\left[ \int d^3 r  \ |\chi_\mu (\bm{r})|^2 \right]^{-1/2}$ is the Gaussian normalization, and $i_q=t,u,v$ for $q=x,y,z$. We, then, expand the electron density in the Gaussian basis
\begin{equation}\label{eqn:n}
n (\bm{r})=\sum_{\mu \nu} P_{\mu\nu} \ \chi_\mu (\bm{r}) \chi_\nu (\bm{r})
\end{equation}
where $P_{\mu\nu}$ are one-electron reduced density matrix coefficients. Using Eqs. \eqref{eqn:mom}-\eqref{eqn:n}, the multipole moments may be expressed as
\begin{widetext}
\begin{equation}\label{eqn:int_multi}
\eta^m_{\ell} \left[ n \right](\mathbf{R})=  \sum_{\mu \nu} N_\mu \ N_\nu \ P_{\mu\nu} \ K_{\mu \nu}  \sum_{tuv}^{\ell}\sum_{t' u' v'}^{\ell'} \sum_{t'' u'' v''}^{\ell''}   D^{\ell m}_{tuv} \ D^{\ell' m'}_{t'u'v'} \ D^{\ell'' m''}_{t''u''v''} \prod_{q=x,y,z}  \int d q  \ (q- R_{q} )^{i_q} (q- R_{\mu,q} )^{i_q'} (q- R_{\nu,q} )^{i_q''} \ e^{- \lambda_{\mu \nu} (q- \Lambda_{\mu \nu,q} )^2}  
\end{equation}
\end{widetext}
where the quantities $\lambda_{\mu \nu}=\alpha_\mu+\alpha_\nu$, while $K_{\mu \nu}={\rm exp}\left[ -\alpha_\mu \alpha_\nu (\mathbf{R}_\mu - \mathbf{R}_\nu)^2 / \lambda_{\mu \nu}  \right]$ and $\boldsymbol{\Lambda}_{\mu \nu}=(\alpha_\mu \mathbf{R}_\mu+ \alpha_\nu \mathbf{R}_\nu) / \lambda_{\mu \nu}$ were derived using the Gaussian product theorem. The remaining integral in Eq. \eqref{eqn:int_multi} can be straightforwardly calculated using the binomial theorem as well as the integral representation of the Hermite polynomial\cite{bell,grad_einstein}
\begin{equation}
H_n (iR) = \frac{(2i)^n}{\sqrt{\pi}} \int dx \ x^n e^{-(x-R)^2}
\end{equation}
To summarize, from Eq. \eqref{eqn:Phi} and \eqref{eqn:int_multi}, we can expand the Hartree potential to arbitrarily high multipoles, by making use of the spherical-to-Cartesian transfer coefficients $D^{\ell m}_{tuv}$ of Eqs. \eqref{eq19} and \eqref{eqD1}.

We apply the multipole procedure to reproducing the Hartree potential due to the molecular Hartree-Fock electron density. To test the implementation, also on high-angular momenta basis functions, we study the La$_2$ diatomic molecule. The La$_2$ molecule, incorporated in a fullerene cage, has generated interest for applications on organic field-effect transistors.\cite{kobayashi2003conductivity,la2_fullerene,zhao2017quantum} We employed the experimental La-La distance of 3.84 \AA,\cite{la2_distance} as well as the Gaussian basis set due to Towler.\cite{towler_basis}

\begin{figure}[ht!!]
\centering
\includegraphics[width=8.6cm]{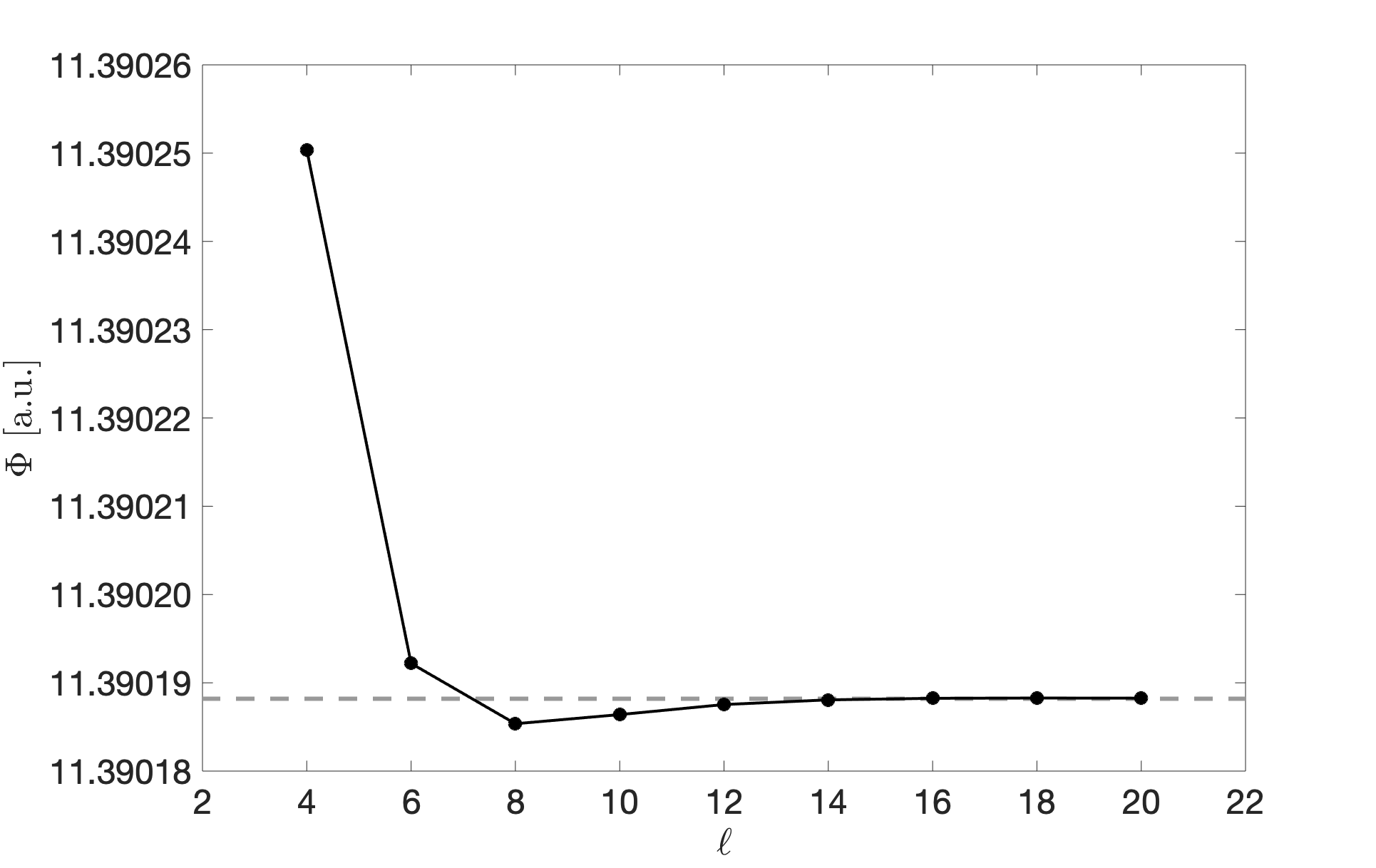}
\caption{Target (dotted line) and multipole (solid line) Hartree potentials $\Phi[n](\bm{r})$ of the La$_2$ molecule at $\bm{r}$ placed 4 \AA $\,$ away from the nearest La nucleus.}
\label{fig:plottino}
\end{figure}

Numerical values of the Hartree potential $\Phi\left[ n\right] (\bm{r})$ (computed through the multipole expansion) with $\bm{r}$ placed 4 \AA $\,$away from the nearest La center, are provided in Table \ref{tab:my_label} and Fig. \ref{fig:plottino}, and compared to the target result, from the first line of Eq. \eqref{eqn:Phi}, which was computed using the semi-analytical Boys-function based method of Ref. \onlinecite{saunders1983molecular}. The comparison shows that the multipole series is visibly convergent, although highly accurate calculations require up to $\ell=14$ multipole moments.

\begin{table}[htb]
    \centering
    \begin{tabular}{c|c}
           $\ell$ & $\Phi[n](\bm{r})$ [a.u.] \\
           \toprule
           4     &      11.3902503651666  \\
           6     &      11.3901921981578  \\
           8     &      11.3901853622202  \\
           10    &      11.3901863980079  \\
           12    &      11.3901875365295  \\
           14    &      11.3901880678308  \\
           16    &      11.3901882440467  \\
           18    &      11.3901882741596  \\
           20    &      11.3901882578498  \\
           \toprule
           $\infty$ &      11.3901881899539 \\
           \toprule
    \end{tabular}
    \caption{Hartree potential $\Phi[n](\bm{r})$ obtained using different values of $\ell$ quantum numbers as upper bound in the summation \eqref{eqn:Phi}. The target value, obtained using the first expression in \eqref{eqn:Phi} through semi-analytical Boys-function based method, is reported in the last line.}
    \label{tab:my_label}
\end{table}

\section{Conclusions}
A formula for the transformation coefficients of complex and real spherical solid harmonics in terms of (exactly factorizable) Cartesian functions has been given. Explicit expressions of the coefficients are tabulated up to the quantum number $\ell=10$. The source code, used to generate the transformation coefficients up to arbitrary $\ell$ quantum numbers, is also provided in the Supporting Information, as well as freely accessible on GitHub.\cite{code_github_link} The coefficients are applied to computing the Hartree potential by an arbitrary-order multipole expansion. The multipole series, although convergent, required up to $\ell=14$ multipole moments for a highly accurate calculation in the case of the La$_2$ molecule.


\begin{thebibliography}{47}
\bibitem [{The general formula for the number of CGTOs at given $\ell$ is
  provided by the ``stars and bars'' (also known as ``sticks and stones'')
  problem of combinatorics and reads $\displaystyle \binom{\ell+2}{2}$ , while
  the number of SGTOs is just $2\ell+1$()}]{notebin}%
  \BibitemOpen
  The general formula for the number of CGTOs at given $\ell$ is provided by
  the ``stars and bars'' (also known as ``sticks and stones'') problem of
  combinatorics and reads $\displaystyle \binom{\ell+2}{2}$ , while the number
  of SGTOs is just $2\ell+1$,\ \href@noop {} {}\BibitemShut {NoStop}%
\end{thebibliography}

%

\end{document}